# SOLAR MAGNETIC FIELDS AND TERRESTRIAL CLIMATE

**Georgieva K.[1], Nagovitsyn Yu.[2], Kirov B.[1]**

*1 - Space and Solar-Terrestrial Research Institute of Bulgarian Academy of Sciences, Sofia, Bulgaria*
*2 - Central Astronomical Observatory at Pulkovo, St.Petersburg, Russia*

## СОЛНЕЧНЫЕ МАГНИТНЫЕ ПОЛЯ И ЗЕМНОЙ КЛИМАТ

**Георгиева К.[1], Наговицын Ю.А.[2], Киров Б.[1]**

*1 – Институт космических исследований и технологий Болгарской Академии Наук, София, Болгария*
*2 – Главная Астрономическая Обсерватория, Санкт Петербург, Россия*

*Солнечная радиация считается одним из основных естественных факторов, влияющих на земной климат, и ее вариации включаются в большинство численных моделей, оценивающих эффекты естественных по сравнению с антропогенными факторами изменений климата. Солнечный ветер, вызывающий геомагнитные возмущения, является другим агентом солнечной активности, чья роль в изменениях климата еще не полностью понята, но активно исследуется. Для целей климатического моделирования важно оценить и прошлые, и будущие вариации солнечной радиации и геомагнитной активности, которые тесно связаны с вариациями солнечных магнитных полей. Прямые измерения солнечных магнитных полей имеются за ограниченный период, но их можно восстановить из измерений геомагнитной активности. Мы представляем реконструкцию общей солнечной радиации, основанной на геомагнитных данных, и прогноз будущей радиации и геомагнитной активности, которые можно включить в модели ожидаемых климатических изменений.*

### 1. Introduction

Knowing the evolution of solar activity is important for both evaluating the past climate variability and how much of it is due to solar variability and how much to human activity, and for estimating the expected variations of solar activity and their eventual response in terrestrial climate.

There are two main geoeffective manifestations of solar activity whose variability is important for the Earth's climate: solar electromagnetic radiation and solar corpuscular emissions. In the present paper we are dealing with the long-term variations in solar electromagnetic radiation (the solar irradiance).

Sun emits electromagnetic radiation in a wide range of wavelengths - from extreme ultraviolet to infrared. The integrated energy entering the terrestrial atmosphere, measured in W/m², is referred to as Total Solar Irradiance (TSI).

For quite a long time, TSI has been regarded as constant, and has been even known as "the solar constant".

## 2. TSI measurements

Space-borne measurements of TSI began in 1978 with the Hickey–Frieden radiometer [1] aboard the NOAA/NASA mission Nimbus-7, followed by the ACRIM experiment aboard the Solar Maximum Mission. The most important result from them is that the "solar constant" is not constant but varies on different time scales, from minutes to decades, being maximum in sunspot maximum and minimum in sunspot minimum [2]. Though sunspots are dark areas on the solar surface and cause reduction of TSI on day-to-day basis, the variations of their number and area in the ~11-year solar cycle are accompanied by variations of the number and area of bright faculae which more than compensate the darkening due to sunspots, and as a result TSI is higher in sunspot maximum than in sunspot minimum [3].

TSI has been measured by a number of instruments covering partly overlapping periods. They all show the solar cycle variations in TSI, but their results are quite different in terms of both absolute levels and trends. Even for one single instrument, the trend may be not reliable because of instrument-related factors like calibration, sensor degradation with time, etc. Therefore, combining the various data sets into a single time series is quite difficult and controversial. The basic disagreement among the published composites [4-6] important for evaluating solar contributions to climate change, is in long-term trends: some composites show decrease in TSI from cycle to cycle, others show increase, or no significant trend.

The differences are even bigger in estimates of TSI over longer periods with no instrumental measurements when proxies are used to reconstruct the irradiance. Reconstructions of TSI since the Maunder minimum in the second half of the 17$^{th}$ century, when the Sun was extremely inactive and Europe experienced the "Little Ice Age", estimate values of TSI from equal to the ones during the last 2008-2009 solar minimum [7, 8] to almost 6 W/m$^2$ lower [9].

The basic assumption in reconstructing variations of TSI is that all changes in the solar irradiance on time scales longer than hours are entirely caused by changes in the solar surface magnetic flux which can be traced through surface features, such as spots, faculae, ephemeral regions [10]. TSI is then determined by the darkening due to sunspots, plus brightening due to faculae and ephemeral regions, plus the contribution of the "quiet Sun" considered constant. Information on the surface coverage by each component and its evolution in time is best provided by direct measurements of the solar photospheric magnetic fields, i.e. by the full-disc magnetograms, but they are also available for a limited period – less than 40 years. For earlier times, records of sunspot number and areas are used. There are no long records of ephemeral areas, and sunspot and facular areas have been measured only since 1874 so before that all their fluxes are estimated based on empirically found ratios with the sunspot number for a recent period when all values are available from measurements [11].

### 3. Problems with TSI reconstructions

Several wrong assumptions are made when reconstructing TSI from only the number of sunspots. The first one is that the sunspot number can be used instead of the sunspot area which actually determines the impact of sunspots reducing the TSI. Actually, the relation between sunspot number and area changes in time, and recently it was shown that the proportion of small to large spots has been increasing [12] which means decreasing ratio of the total sunspot area to the number of sunspots – Fig.1.

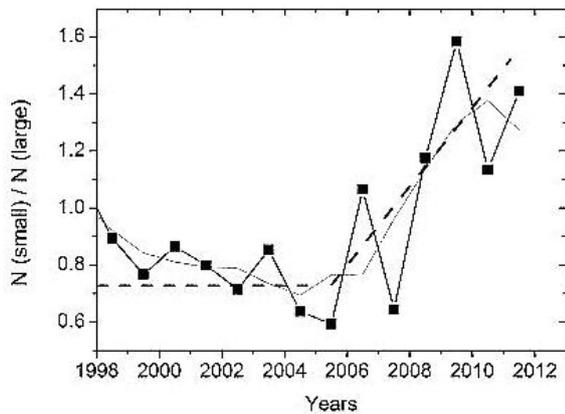

Second, the darkness of a sunspot important for its contribution to TSI depends on its magnetic field [13] which in turn is related to the spot's area [14]. However, the relation between the spots' area and magnetic field changes from cycle to cycle – Fig.2.

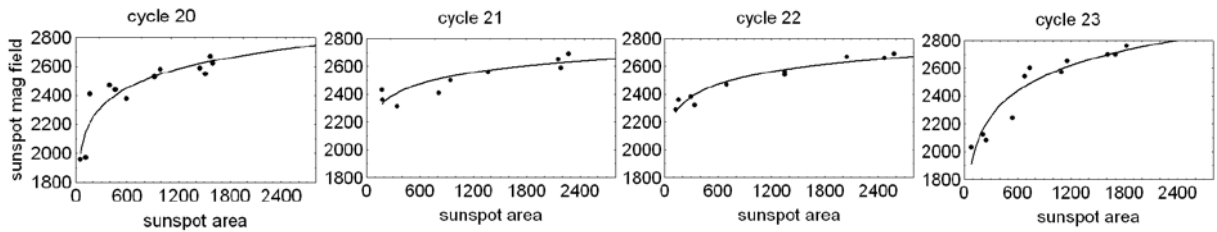

Moreover, this nonlinear relation itself depends on the sunspots' magnetic field (Fig.3), so it is not correct to estimate by how much a number of sunspots decrease TSI without information about their magnetic fields. The average sunspot magnetic field in cycle 20-23 used in Fig.3 is from a composite historical synoptic data set described in [15].

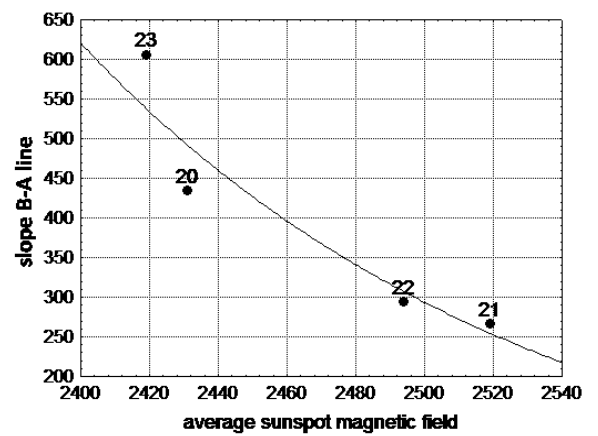

The most important problem is the estimation of the area of bright faculae whose contribution to TSI is calculated to be about 50% higher than the one of sunspots [16]. In reconstructions of TSI for longer periods [11], the facular area value is based on the correlation found between the facular and sunspots areas in cycle 22 [16]. However, this correlation is not constant either. It varies with solar magnetic fields as the cycle progresses [11] and from cycle to cycle [17,

18]. As a result, the amplitudes of the 11-year solar cycles as measured by the facular and sunspots areas are not well correlated [11].

Therefore, to be able to correctly reconstruct TSI, we need information about the magnitude and evolution of solar magnetic fields. There are no long-term records of solar magnetic fields, but we have found that geomagnetic records reflect the variations of the sunspot magnetic fields: the geomagnetic activity "floor" (the value below which geomagnetic activity cannot fall even in the lack of any sunspots) in a sunspot cycle is proportional to the sunspot magnetic field in the sunspot cycle minimum, and the rate of increase of geomagnetic activity with increasing number of sunspots is proportional to the rate of increase of sunspot magnetic field from cycle minimum to cycle maximum [19]. These correlations can be used to reconstruct the solar magnetic fields and to account for their variations in reconstructions of long term (cycle to cycle) variations of total solar irradiance. Here we use the *aa*-index of geomagnetic activity and the international sunspot number for the period since 1868, and the ESAI data base of Extended Solar Activity Indices [20] for earlier periods to estimate the solar magnetic fields as described in [19], and we regress them to the ACRIM [4] composite TSI series to derive the correlations between sunspot magnetic fields and TSI, and to estimate the long term (cycle to cycle) variations of TSI.

## 4. Reconstructions from geomagnetic data

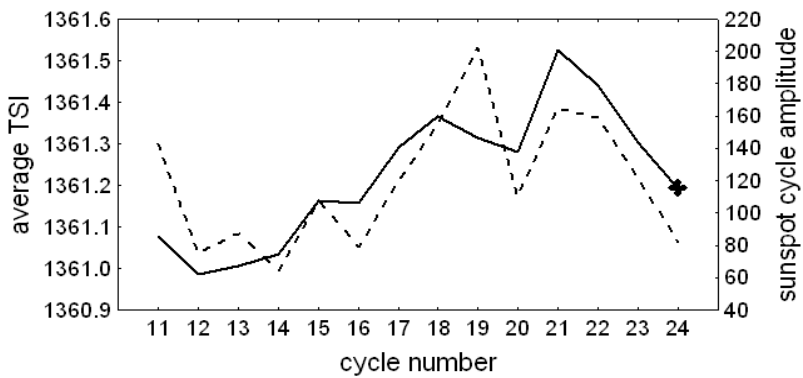

The solid line in Fig.3 presents the calculated TSI from cycle 11 (1867-1877) to cycle 23 (1996-2007). Each point is the TSI averaged over one whole cycle. A part of cycle 24 (2008-2013) is also included for comparison but it should be kept in mind that as the cycle is not yet complete and this period does not cover the declining phase of the cycle, the value for the full cycle will be lower. The dashed line presents the amplitude (maximum sunspot number) in the respective cycles.

A couple of peculiarities can be seen in Fig.3. First, the TSI averaged over the solar cycle has increased by 0.6 W/m$^2$ from a minimum value in cycle 12 (1878-1890) to a maximum value in cycle 21 (1986-1995). The global temperature response to changing TSI is $\Delta T = \lambda \cdot \Delta TSI$ with $\Delta TSI$ in W/m$^2$, and different estimated put $\lambda$ (climate sensitivity to TSI change) in the range 0.3-1.8 K/Wm$^{-2}$ which means that TSI variations are responsible for between 0.2 and 1.1 $^o$K change in global temperature during the last century.

Next, the variations in TSI follow roughly the variations in the sunspot cycles amplitudes, however with important differences. The most intense cycle in this period as measured by the number of sunspots was cycle 19, but the most intense cycle as measured by TSI was cycle 21. Similarly, the weakest cycle in sunspot number was cycle 14, but the weakest cycle in TSI was cycle 12. The explanation for these discrepancies is that the cycle averaged TSI is a result of the interplay between the variations of the darkening determined by the total area and magnetic field in sunspots, and the brightening determined by the total area and magnetic field in facular and ephemeral regions. As pointed out in [16-18], the ratio between the two quantities varies, and the maximum in the area of faculae which is about 50% more important for TSI than the sunspot area, does not coincide with the sunspot area maximum. As a result, the increase in TSI lasted until cycle 21 (1976-1985), unlike sunspot number which began decreasing about 20 years earlier after a maximum in cycle 19.

From the ESAI database of solar and geomagnetic activity, we can also reconstruct sunspot magnetic fields, and estimate the TSI since the beginning of the 17$^{th}$ century (Fig. 4). Since the end of the Maunder minimum (cycle -4, 1698 -1712), TSI has increased by about 3 W/m$^2$, and since the deepest part of the Maunder minimum (not shown), the increase is about 7 W/m$^2$, in good agreement with the results of [9] and much more than the estimations of [21].

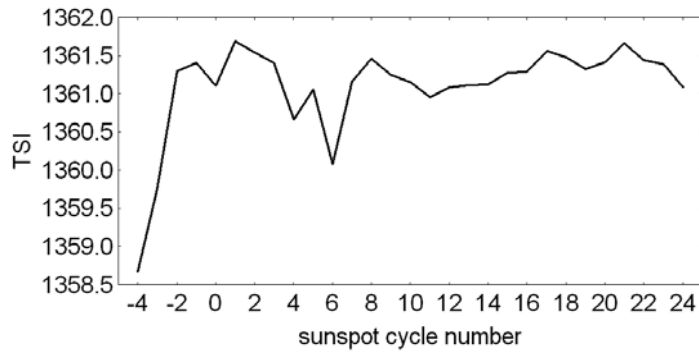

### 5. Conclusion

Total solar irradiance reconstructions, calculated taking into account the evolution of sunspot magnetic fields derived from geomagnetic data, support the TSI composites and reconstructions showing much higher long-term TSI variability, and consequently much bigger solar influences on climate variability than accounted for in popular models. Even for a very conservative value of climate sensitivity to TSI variations of 0.5 adopted by IPCC AR4, the estimated TSI increase since the early 18$^{th}$ century (the end of the Maunder minimum), and moreover since the deepest part of the Maunder minimum, demonstrate that TSI increase alone was responsible for ΔT of at least 1.5 and 3.5 $^{o}$K, respectively.

It should be noted that solar electromagnetic radiation is only one of the solar agents affecting climate. Another one is the solar wind – the ever expanding solar corona filling the whole heliosphere with solar plasma and embedded magnetic fields. Its role in climate change is a subject of extensive research, but is not yet fully understood, and its contribution is included in only a few atmospheric models. But it has been found that Earth's surface temperature is correlated with both decadal averages and solar cycle minimum values of the geo-magnetic activity [22] which are determined by the solar wind parameters [23]. The solar cycle minimum values of the geomagnetic activity $aa_{min}$ (geomagnetic activity floor) and $b$ (rate of increase of geomagnetic activity with increasing sunspot numbers) both have secular periodic variations (Fig.5) which make it possible to forecast their future variations. The expected decrease in $aa_{min}$ means decrease in surface temperature in the next decades. On the other hand, $aa_{min}$ is proportional to sunspot magnetic field in cycle minimum, which is positively correlated with cycle averaged TSI, and $b$ is proportional to the rate of increase of sunspot magnetic fields with increasing sunspot number which is negatively correlated to TSI. Therefore, the expected decrease of $aa_{min}$ and increase of $b$ both predict future decrease of TSI which, added to the expected decreasing geomagnetic activity, will be an additional factor for the future global cooling.

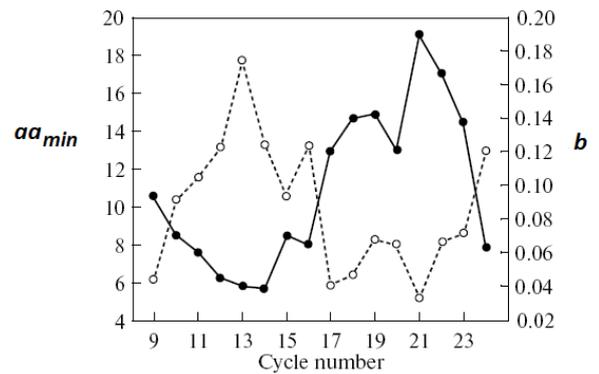

**Acknowledgements:** This work is done as a part of a joint research project of RAS and BAS. The idea evolved from a workshop funded by COST ES1005 project TOSCA.